\documentclass[aps,prb,preprint,showpacs,superscriptaddress]{revtex4}
\usepackage{bm, amsmath, amssymb}
\usepackage{graphicx,subfigure}
\newcommand{\figwidth}{0.47\textwidth}

\begin{document}

\title{Inertial longitudinal magnetization reversal for
non-Heisenberg ferromagnets }

\author{E.~G. Galkina}
\affiliation{Institute of Physics, 03028 Kiev, Ukraine}
\affiliation{CEMS, RIKEN, Saitama, 351-0198, Japan}

\author{V.~I. Butrim}
\affiliation{Vernadsky Taurida National University, Simferopol,
95007 Ukraine }

\author{Yu.~A. Fridman}
\affiliation{Vernadsky Taurida National University, Simferopol,
95007 Ukraine }

\author{B.~A. Ivanov}
\email{bivanov@i.com.ua} \affiliation{Institute of Magnetism, 03142,
Kiev, Ukraine} \affiliation{Taras Shevchenko National University of
Kiev, 01601, Ukraine} \affiliation{CEMS, RIKEN, Saitama, 351-0198,
Japan}

\author{Franco Nori}
 \affiliation{CEMS, RIKEN, Saitama, 351-0198, Japan}
 \affiliation{Department of Physics, The University
of Michigan, Ann Arbor, MI 48109-1040, USA.}

\date{\today}

\begin{abstract}
We analyze theoretically the novel pathway of ultrafast spin
dynamics for ferromagnets with high enough single-ion anisotropy
(non-Heisenberg ferromagnets). This longitudinal spin dynamics
includes the coupled oscillations of the modulus of the
magnetization together with the quadrupolar spin variables, which
are expressed through quantum expectation values of operators
bilinear on the spin components. Even for a simple single-element
ferromagnet, such a dynamics can lead to an inertial magnetization
reversal under the action of an ultrashort laser pulse.
\end{abstract}


% 75.10.Jm quantum spin models
% 75.10.Hk : Classical spin models
% 78.47.J- Ultrafast pump/probe spectroscopy
% 05.45.-a Nonlinear dynamics and nonlinear dynamical systems.

\date{\today}

 \maketitle

\section{Introduction}
\label{sec:intro}

Which is the fastest way to reverse the magnetization of either a
magnetic particle or a small region of a magnetic film? This
question has attracted significant interest, both fundamental and
practical, for magnetic information storage.~\cite{[1]} Intense
laser pulses, with durations less than a hundred femtoseconds, are
able to excite the ultrafast evolution of the total spin of a
magnetically-ordered system on a picosecond time scale, see e.g. the
reviews.~\cite{[2],[3],[4]} The limitations for the time of
magnetization reversal come from the characteristic features of the
spin evolution for a magnet with a concrete type of magnetic order.
The dynamical time cannot be shorter than the characteristic period
of spin oscillations $T$, $T=2\pi /\omega _0 $, where $\omega _0 $
is the magnetic resonance frequency. For ferromagnets, the frequency
of standard spin oscillations (precession) is $\omega
_{0,\mathrm{FM}} =\gamma H_\mathrm{r} $, where $\gamma $ is the
gyromagnetic ratio, and $H_\mathrm{r} $ is an effective field of
relativistic origin, like the anisotropy field, which is usually
less than a few Tesla. Thus, the dynamical time for Heisenberg
ferromagnets cannot be much shorter than one nanosecond. For
antiferromagnets, all the dynamical characteristics are exchange
enhanced, and $\omega _{0,\mathrm{AFM}} =\gamma \sqrt
{H_{\mathrm{ex}} H_\mathrm{r} } $, where $H_{\mathrm{ex}} $ is the
exchange field, $H_{\mathrm{ex}} =J/2\mu _\mathrm{B} $, $J$ is the
exchange integral, and $\mu _\mathrm{B} $ is the Bohr magneton, see
Ref.~\onlinecite{[5]}. The excitation of terahertz spin oscillations
has been experimentally demonstrated for transparent
antiferromagnets using the inverse Faraday effect or the inverse
Cotton-Mouton effect.~\cite{[6],[7],[8],[9],[10],[11]} The
non-linear regimes of such dynamics include the inertia-driven
dynamical reorientation of spins on a picosecond time scale, which
was observed in orthoferrites.~\cite{[10],[11]}

The exchange interaction is the strongest force in magnetism, and
the exchange field $H_{\mathrm{ex}} $ can be as strong as $10^3$
Tesla. The modulus of the magnetization is determined by the
exchange interaction, and the direction of the magnetization is
governed by relativistic interactions. It would be very tempting to
produce a magnetization reversal by changing the modulus of the
magnetization vector, i.e., via the \textit{longitudinal} dynamics
of ${\rm {\bf M}}$. For such a process, dictated by the exchange
interaction, the characteristic times could be of the order of the
exchange time $\tau _{\mathrm{ex}} =1/\gamma H_{\mathrm{ex}} $,
which is shorter than one picosecond. However, within the standard
approach such dynamics is impossible. The evolution of the modulus
of the magnetization, $M=\vert {\rm {\bf M}}\vert $, within the
closed Landau-Lifshitz equation for the magnetization only (or the
set of such equations for the sublattice magnetizations, ${\rm {\bf
M}}_\alpha )$, is purely dissipative.~\cite{[12]} This feature could
be explained as follows: two angular variables, $\theta $ and
$\varphi $, describing the direction of the vector ${\rm {\bf M}}$,
within the Landau-Lifshitz equation determine the pair of conjugated
Hamilton variables $(\cos \theta $ and $\varphi $ are the momentum
and coordinate, respectively). Also, the evolution of the single
remaining variable $M=\vert {\rm {\bf M}}\vert $, governed by a
first-order equation can be only dissipative; see a more detailed
discussion below. Moreover, the exchange interaction conserves the
total spin of the system, and the relaxation of the total magnetic
moment of any magnet can be present only when accounting for
relativistic effects. Thus the relaxation time for the total
magnetic moment is relativistic but it is exchange-enhanced, as was
demonstrated within the irreversible thermodynamics of the magnon
gas.~\cite{[12]} Note here that the relaxation of the magnetization
of a \textit{single sublattice} for multi-sublattice magnets can be
of purely exchange origin.~\cite{[13]} Recently, magnetization
reversal on a picosecond time scale has been experimentally
demonstrated for the ferrimagnetic alloy GdFeCo, see
Refs.~\onlinecite{[13],[14]}. These results can be explained within
the concept of exchange relaxation, developed by
Baryakhtar,~\cite{[15]} accounting for the purely exchange evolution
of the sublattice magnetization.~\cite{[16]} Such an exchange
relaxation can be quite fast, but its characteristic time is again
longer than the expected ``exchange time'' $\tau _{\mathrm{ex}}
=1/\gamma H_{\mathrm{ex}} $.

Thus, the ultrafast mechanisms of magnetization reversal implemented
so far are: the dynamical (inertial) switching possible for
antiferromagnets,~\cite{[10],[11]} and the exchange longitudinal
evolution for ferrimagnets.~\cite{[13],[14],[16]} These are both
quite fast, with a characteristic time of the order of picoseconds;
but their characteristic times are longer than the ``ideal
estimate'': the exchange time $\tau _{\mathrm{ex}} $.

In this work, we present a theoretical study of the possibility of
the \textit{dynamical} evolution of the modulus of the magnetization
for non-Heisenberg ferromagnets with high enough single-ion
anisotropy that can be called \textit{longitudinal spin dynamics}.
For such a dynamics, an \textit{inertial magnetization reversal} is
possible even for a simple single-element ferromagnet. Longitudinal
dynamics does not exist in Heisenberg magnets, and this dynamics
cannot be described in terms of the Landau-Lifshitz equation, or
using the Heisenberg Hamiltonian, which is bilinear over the
components of spin operators for different spins, see more details
below in Sec. II. The key ingredient of our theory is the inclusion
of higher-order spin quadrupole variables. It is known that for
magnets with atomic spin $S>1/2$, allowing the presence of
single-ion anisotropy, the spin dynamics is not described by a
closed equation for spin \textit{dipolar} variable $\left\langle
{\rm {\bf S}} \right\rangle $ alone (or magnetization ${\rm {\bf
M}}=-2\mu _\mathrm{B} \left\langle {\rm {\bf S}} \right\rangle
$).~\cite{[17],[18],[19],[20],[21],Chubuk90,Matveev,AndrGrishchuk}
Here and below $\left\langle {...} \right\rangle $ means quantum and
(at finite temperature) thermal averaging. To be specific, we choose
the spin-one ferromagnet with single-ion anisotropy, the simplest
system allowing this effect. The full description of these magnets
requires taking into account the dynamics of \textit{quadrupolar
}variables, $S_{ik} =(1/2)\left\langle {S_i S_k +S_k S_i }
\right\rangle $, that represent the quantum averages of the
operators, bilinear in the spin components. Our theory is based on
the consistent semiclassical description of a full set of spin
quantum expectation values (dipolar and quadrupolar) for the
spin-one system, which was investigated by many authors from
different
viewpoints.~\cite{[17],[18],[19],[20],[21],Chubuk90,Matveev,AndrGrishchuk}
As we will show, the longitudinal dynamics of spin, including
nonlinear regimes, can be excited by a femtosecond laser pulse. With
natural accounting for the dissipation, the longitudinal spin
dynamics can lead to changing the sign of the total spin of the
system (longitudinal magnetization reversal).

\section{ MODEL DESCRIPTION}
\label{sec:model} The Landau-Lifshitz equation was proposed many
years ago as a phenomenological equation, and it is widely used for
the description of various properties of ferromagnets. Concerning
its quantum and microscopic basis, it is worth noting that this
equation naturally arises using the so-called \textit{spin coherent
states}.~\cite{[22],[23]} These states can be introduced for any
spin $S$ as the state with the maximum value of the spin projection
on an arbitrary axis ${\rm {\bf n}}$. Such states can be
parameterized by a unit vector ${\rm {\bf n}}$; the direction of the
latter coincides with the quantum mean values for the spin operator
$\left\langle {\rm {\bf S}} \right\rangle =S{\rm {\bf n}}$ (dipolar
variables). This property is quite convenient for linking the
quantum physics of spins to a phenomenological Landau-Lifshitz
equation. The use of spin coherent states is most efficient when the
Hamiltonian of the system is linear with respect to the operators of
the spin components. If an initial state is described by a certain
spin coherent state, its quantum evolution will reduce to a
variation of the parameters of the state (namely, the direction of
the unit vector ${\rm {\bf n}})$, which are described by the
classical Landau-Lifshitz equation. Thus, spin coherent states are a
convenient tool for the analysis of spin Hamiltonians containing
only operators linear on the spin components or their products on
different sites. An important example is the bilinear Heisenberg
exchange interaction, described by the first term in Eq. (\ref{eq1})
below.

In contrast to the cases above, for the full description of spin-$S$
states, one needs to introduce $SU(2S+1)$ generalized coherent
states.~\cite{[17],[19],[20],[21]} The analysis shows that spin
coherent states are less natural for the description of magnets
whose Hamiltonian contains products of the spin component operators
at a single site. Such terms are present for magnets with single-ion
anisotropy or a biquadratic exchange interaction. Magnets with
non-small interaction of this type are often called
\textit{non-Heisenberg}. For such magnets, some non-trivial
features, absent for Heisenberg magnets, are known. Among them we
note the possibility of so-called quantum spin reduction; namely the
possibility to have the value of $\vert \left\langle {\rm {\bf S}}
\right\rangle \vert $ less than its nominal value, $\vert
\left\langle {\rm {\bf S}} \right\rangle \vert <S$, even for pure
states at zero temperature. This was first mentioned by
Moriya,~\cite{[24]} as early as 1960. As an extreme realization of
the effect of quantum spin reduction, we note the existence of the
so-called spin nematic phases with a zero mean value of the spin in
the ground state at zero temperature. In the last two decades, the
interest on such states has been considerable, motivated by studies
of multicomponent Bose-Einstein condensates of atoms with non-zero
spin.~\cite{[25],[26],[27],[28]}

A significant manifestation of quantum spin reduction is the
appearance of an additional branch of the spin oscillations, which
is characterized by the dynamics (oscillations) of the length of the
mean value of spin without spin
precession.~\cite{[17],[18],[19],[20],[21],Chubuk90,Matveev} The
characteristic frequency of this mode can be quite high (of the
order of the exchange integral). For this reason, for a description
of resonance properties or thermodynamic behavior of magnets, this
mode is usually neglected, and the common impression is that the
dynamics of magnetic materials with constant single-ion anisotropy
$K<$(0.2-0.3)$J$ is fully described by the standard phenomenological
theory. However, for an ultrafast evolution of the spin system under
a femtosecond laser pulse, one can expect a lively demonstration of
this longitudinal high-frequency mode. Thus, it is important to
explore the possible manifestations of the effects of quantum spin
reduction in the dynamic properties of ferromagnets.

The simplest model allowing spin dynamics with effects of quantum spin
reduction is described by the Hamiltonian
\begin{equation}
\label{eq1} H=-\frac{1}{2}\sum\limits_{\mathbf{n},\boldsymbol{\ell}
} {\bar {J}{\rm {\bf S}}_\mathbf{n} {\rm {\bf
S}}_{\mathbf{n}+\boldsymbol{\ell} } }
+\frac{K}{2}\sum\limits_\mathbf{n} {\left( {S_{\mathbf{n},x} }
\right)^2} ,
\end{equation}
where ${\rm {\bf S}}_\mathbf{n} $ is the spin-one operator at the
site $\mathbf{n}$; $\bar {J}>0$ is the exchange constant for
nearest-neighbors $\boldsymbol{\ell} $, and $K>0$ is the constant of
the easy-plane anisotropy with the plane $yz$ as the easy plane. The
quantization axis can be chosen parallel to the $z$-axis and
$\left\langle {\rm {\bf S}} \right\rangle =\left\langle {S_z }
\right\rangle {\rm {\bf e}}_z $. For the full description of spin
$S=1$ states, let us introduce $SU(3)$ coherent
states~\cite{[17],[19],[20],[21]}
\begin{equation}
\label{eq2} \left| {\mathbf{u},\mathbf{v}} \right\rangle
=\sum\limits_{j=x,y,z} {\left( {u_j +iv_j } \right)\left| {\psi _j }
\right\rangle } ,
\end{equation}
where the states $\left| {\psi _j } \right\rangle $ determine the
Cartesian states for $S=1$ and are expressed in terms of the
ordinary states \{$\left| {\pm 1} \right\rangle ,\;\left| 0
\right\rangle $\} with given projections $\pm 1,\;0$ of the operator
$S_z $ by means of the relations $\left| {\psi _x } \right\rangle
=\left( {\left| {-1} \right\rangle -\left| {+1} \right\rangle }
\right)/\sqrt 2 $, $\left| {\psi _y } \right\rangle =i\left( {\left|
{-1} \right\rangle +\left| {+1} \right\rangle } \right)/\sqrt 2 $,
$\left| {\psi _z } \right\rangle =\left| 0 \right\rangle $, with the
real vectors $\mathbf{u}$ and $\mathbf{v}$ subject to the
constraints $\mathbf{u}^2+\mathbf{v}^2=1$, $\mathbf{u}\cdot
\mathbf{v}=0$. All irreducible spin averages, which include the
dipolar variable $\left\langle \mathbf{S} \right\rangle $ (average
value of the spin) and quadrupole averages $S_{ik} $, bilinear over
the spin components, can be written through $\mathbf{u}$ and
$\mathbf{v}$ as follows
\begin{eqnarray}
\label{eq3} \left\langle \mathbf{S} \right\rangle &=&2(\mathbf{u}\times \mathbf{v}), \nonumber \\
 S_{ik}
=\frac{1}{2}\left\langle {S_i S_k +S_k S_i } \right\rangle &=&\delta
_{ik} -u_i u_k -v_i v_k .
\end{eqnarray}
At zero temperature and within the mean-field approximation, the
spin dynamics is described by the Lagrangian~\cite{[21]}
\begin{equation}
\label{eq4} L=-2\hbar \sum\limits_n {\mathbf{v}_n (\partial
\mathbf{u}_n /\partial t} )-W\left( {\mathbf{u},\mathbf{v}} \right),
\end{equation}
where $W\left( {\mathbf{u},\mathbf{v}} \right)=\left\langle
{\mathbf{u},\mathbf{v}} \right|H\left| {\mathbf{u},\mathbf{v}}
\right\rangle $ is the energy of the system.

We are interested in spin oscillations which are uniform in space,
and hence we assume that the discrete variables $\mathbf{u}$ and
$\mathbf{v}$ have the same values for all spins and are only
dependent on time. The frequency spectrum of linear excitations,
which consists of two branches, can be easily obtained on the basis
of the linearized version of the Lagrangian (\ref{eq4}). In the
general case, the system of independent equations for $\mathbf{u}$
and $\mathbf{v}$, taking into account the aforementioned constrains
$\mathbf{u}^2+\mathbf{v}^2=1$, $\mathbf{u}\cdot \mathbf{v}=0$,
consists of four nonlinear equations, describing two different
regimes of spin dynamics. One regime is similar to that for an
ordinary spin dynamics treated on the basis of the Landau-Lifshitz
equation; it corresponds to oscillations of the spin direction. The
second regime corresponds to oscillations of the modulus of the
magnetization $\left\langle {\rm {\bf S}} \right\rangle =S(t){\rm
{\bf e}}_z $, with the vectors $\mathbf{u}$ and $\mathbf{v}$
rotating in the $xy$-plane perpendicular to $\left\langle {\rm {\bf
S}} \right\rangle $. This mode of the spin oscillations corresponds
to the longitudinal spin dynamics. It is convenient to consider
these two types of dynamics separately. Particular non-linear
longitudinal solutions, with $\left\langle {\rm {\bf S}}
\right\rangle =s(t){\rm {\bf e}}_z $ and $u_z =0,\;v_z =0$, were
found in Refs.~\onlinecite{[29],[30]}. Note here that the
longitudinal dynamics is much faster than the standard transversal
one, and the standard spin precession (described by the
Landau-Lifshitz equation) at a picosecond time scale just cannot
develop. Therefore, these two regimes, longitudinal and transverse,
can be treated independently, and we limit ourselves only to the
longitudinal dynamics with $\left\langle {\rm {\bf S}} \right\rangle
=s(t){\rm {\bf e}}_z $ and $u_z =0,\;v_z =0$.

\section{LONGITUDINAL SPIN DYNAMICS}
\label{sec:dynam}

To describe the longitudinal spin dynamics, it is convenient to
introduce new variables: the spin modulus $s=2\vert {\rm {\bf
u}}\vert \vert {\rm {\bf v}}\vert =2uv$ and angular variable $\gamma
$, with
\begin{equation}
\label{eq5} {\rm {\bf u}}=u\left( {{\rm {\bf e}}_x \cos \gamma -{\rm
{\bf e}}_y \sin \gamma } \right),\;{\rm {\bf v}}=v\left( {{\rm {\bf
e}}_x \sin \gamma +{\rm {\bf e}}_y \cos \gamma } \right),
\end{equation}
In this representation $\left\langle {S_z } \right\rangle =s$, and
the non-trivial quadrupolar variables are $\left\langle {S_x S_y
+S_y S_x } \right\rangle =\sqrt {1-s^2} \sin 2\gamma $ and
$\left\langle {S_y^2 -S_x^2 } \right\rangle =\sqrt {1-s^2} \cos
2\gamma $, with all other quantum averages being either zero (as the
transverse spin components $\left\langle {S_{x,y} } \right\rangle $
or $S_{xz} $, $S_{yz} $) or trivial, independent on $s$ and $\gamma
$, as $\left\langle {S_z^2 } \right\rangle =1$. The mean-field
energy, written per one spin through the variables $s,\;\gamma $,
takes the form
\begin{equation}
\label{eq6}
W(s,\gamma )=-\frac{J}{2}s^2-\frac{K}{4}\sqrt {1-s^2} \cos 2\gamma ,
\end{equation}
where $J=\bar {J}Z$, $Z$ is the number of nearest neighbors. The ground
state at $\sqrt {1-s^2} >0$ corresponds to $\cos 2\gamma =1$, with the mean
value of the spin $s=\pm \bar {s}$, $\bar {s}=\sqrt {1-\kappa ^2} <1$, that
is a manifestation of quantum spin reduction at non-zero anisotropy. Here we
introduce the dimensionless parameter $\kappa =K/4J$. For these variables,
the Lagrangian can be written as
\begin{equation}
\label{eq7}
L=\hbar s\frac{\partial \gamma }{\partial t}-W(s,\gamma ),
\end{equation}
and $\hbar s$ and $\gamma $ play the role of canonical momentum and
coordinate, respectively, with the Hamilton function $W(s,\gamma )$.
The physical meaning of the above formal definitions is quite clear:
the angular variable $\gamma $ describes the transformation of
quadrupolar variables under rotation around the $z$-axis, with
$\hbar s$ as the projection of the angular momentum on this axis.

\subsection{Small oscillations}
\label{ssec:small}

Let us now start with the description of the dynamics of
small-amplitude oscillations. After linearization around the ground
state, the equation leads to a simple formula for the frequency of
longitudinal oscillations
\begin{equation}
\label{eq8}
\hbar \omega _l =2J\bar {s}=2J\sqrt {1-\kappa ^2} ,
\end{equation}
which are in fact coupled oscillations of the projection of the spin
and quadrupolar variables, see Fig.~1.

One can see that, for a wide range of values of the anisotropy constant,
like $\kappa <0.2$-0.8, this frequency $\omega _l $ is of the order of
(1.8-1.2)$J/\hbar $, i.e., $\omega _l $ is comparable to the exchange
frequency $J/\hbar $. Thus the longitudinal spin dynamics is expected to be
quite fast. In contrast, standard transversal oscillations for a purely
easy-plane model (\ref{eq1}) are gapless (they acquire a finite gap when accounting
for a magnetic anisotropy in the easy plane, which is usually small). Thus
the essential difference in the frequencies of these two dynamical regimes
is clearly seen.

\begin{figure}%[htbp]
\includegraphics[width=\figwidth]{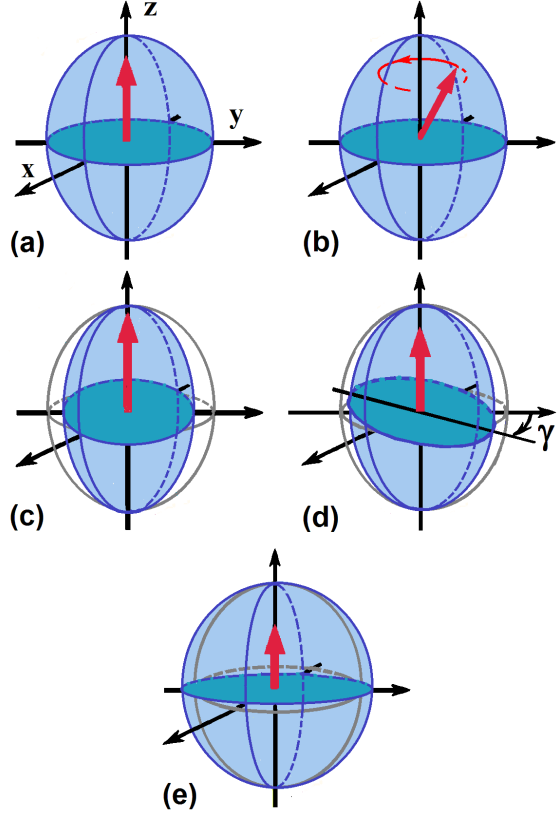}
\caption{\label{fig3}  (color online) Graphic presentation of the
variables $s$ and $\gamma $ and their evolution. The thick red arrow
represents the mean value of the spin. The quadrupolar variables are
shown by the blue three-axial ellipsoid with the directions of the
main axis (chosen to have $\left\langle {S_1 S_2 } \right\rangle
=0)$: ${\rm {\bf e}}_3 ={\rm {\bf e}}_z $ and ${\rm {\bf e}}_{1} $,
${\rm {\bf e}}_{2} $. The half-axes of the ellipsoid are equal to
$\left\langle {S_1 ^2} \right\rangle $, $\left\langle {S_2 ^2}
\right\rangle $ and $\left\langle {S_3 ^2} \right\rangle
=\left\langle {S_z ^2} \right\rangle =1$. (a) the ground state, (b)
the standard transverse dynamics, i.e. the spin precession. The
other frames (c)-(e) present the transient values of the variables
in longitudinal oscillations. (c) and (e) correspond to the longest
and shortest length of the spin, and at the moment depicted in (d)
the spin length equals to its equilibrium value, but the quadrupolar
ellipsoid is turned on the angle $\gamma $ with respect to the
$x$-axis. On the frames (c)-(e), the shape of the unperturbed
ellipsoid is shown by light grey. }
\end{figure}

At a first glance, there is a contradiction between the concept of
longitudinal dynamics caused by single-ion anisotropy and the result
present in equation (\ref{eq8}): the value of $\omega _l $ is still
finite for vanishing anisotropy constant $K$; and it is even growing
to the value $2J/\hbar $ when $\kappa \to 0$. This can be explained
as follows: for a given energy, the ratio of amplitudes for the
oscillations of the spin variable $s$ and quadrupolar variable
$\gamma $ vanish at $\kappa \to 0$ as $\kappa $. In fact, for
extremely low anisotropy the spin oscillations are not present in
this mode, which becomes just a free rotation of the quadrupole
ellipsoid of the form $\gamma =2Jt/\hbar $, with $s=\bar
{s}=\mathrm{const}$. On a phase plane with coordinates $(s,\gamma )$
this dynamics is depicted by vertical straight lines parallel to the
$\gamma $-axis, see Fig. 2(c) below. We will discuss this feature in
more detail with the analysis of non-linear oscillations.

\subsection{Nonlinear dynamics and phase plane analysis.}
\label{ssec:nonlin}

Before considering damped oscillations, it is instructive to discuss
dissipationless non-linear longitudinal oscillations. It is convenient to
present an image of the dynamics as a ``phase portrait'' on the plane
momentum-coordinate $(s,\;\gamma )$, which shows the behavior of the system
for arbitrary initial conditions. The phase trajectories in the plane
without dissipation can be found from the condition $W(s,\gamma
)=\mathrm{const}$.

The energy (\ref{eq6}) has an infinite set of minima, with $s=\pm
\bar {s}$ and $\gamma =\pi n,$ with equal energies (green ellipses
on the Fig.2), and an infinite set of maxima at $s=0$ and $\gamma
=\pi /2+\pi n$ (red ellipses on the Fig.2), here $n$ is an integer.
Only the minima with $s=\bar {s}$ and $s=-\bar {s}$ are physically
different; equivalent extremes with different values of $n$
correspond to the equal values of the observables and are completely
equivalent. The minima on the phase plane correspond to foci with
two physically different equilibrium states with antiparallel
orientation of spin $s=\pm \bar {s}$, and $\left\langle {S_y^2
-S_x^2 } \right\rangle =\sqrt{1-\bar {s}^2}$, $\left\langle S_z^2
\right\rangle =1$,  $\left\langle {S_x S_y +S_y S_x } \right\rangle
=0$. The saddle points are located at the values $s=0$ and $\gamma
=\pi n$. The lines with $s=\pm 1$ are singular; these correspond to
degenerate motion with $\gamma $ linear in time $\gamma =\pm tJ
/\hbar$; the points at these lines where $d\gamma /dt$ change sign,
can be treated as some non-standard saddle points.

The shape of the phase trajectories, i.e., the characteristic
features of oscillations, varies with the change of the anisotropy
parameter $\kappa $. Note first the general trend, the relative
amplitude of the changes of the spin and $\gamma $ depends on
$\kappa $: the bigger $\kappa $ is, the larger values of the change
of spin are observed. The topology of the phase trajectories change
at the critical value of the anisotropy parameter $\kappa $. At
small $\kappa <1/2$, the trajectories with infinite growing $\gamma
$ are present, and the standard separatrix trajectories connect
together different saddle points, see Fig.~2 (c). As mentioned
above; only such trajectories are present at the limit $\kappa \to
0$, but, in fact, even for the small value of $\kappa =0.2$ used in
Fig.2 (c), the main part of the plane is occupied by the
trajectories that change the spin. At the critical value $\kappa
=1/2,$ the separatrix trajectories connect the saddle points at $
s=0 $, $\gamma =0 $  and the degenerated saddle points at the
singular lines with $s=\pm 1$. At larger anisotropy $\kappa
>1/2$, as in Fig. 2(a), the only separatrix loops entering the same saddle
point are present.

The minima, saddle point and the separatrix are the key ingredients
for the switching between the ground states with $s=\bar {s}$ and
$s=-\bar {s}$. The case of large anisotropy $\kappa >1/2$ looks like
the standard one for the switching phenomena, whereas for small
anisotropy the situation is more complicated.

\begin{widetext}

\begin{figure}
 \subfigure[$\ \kappa =$0.6]{\includegraphics[width = 58mm]{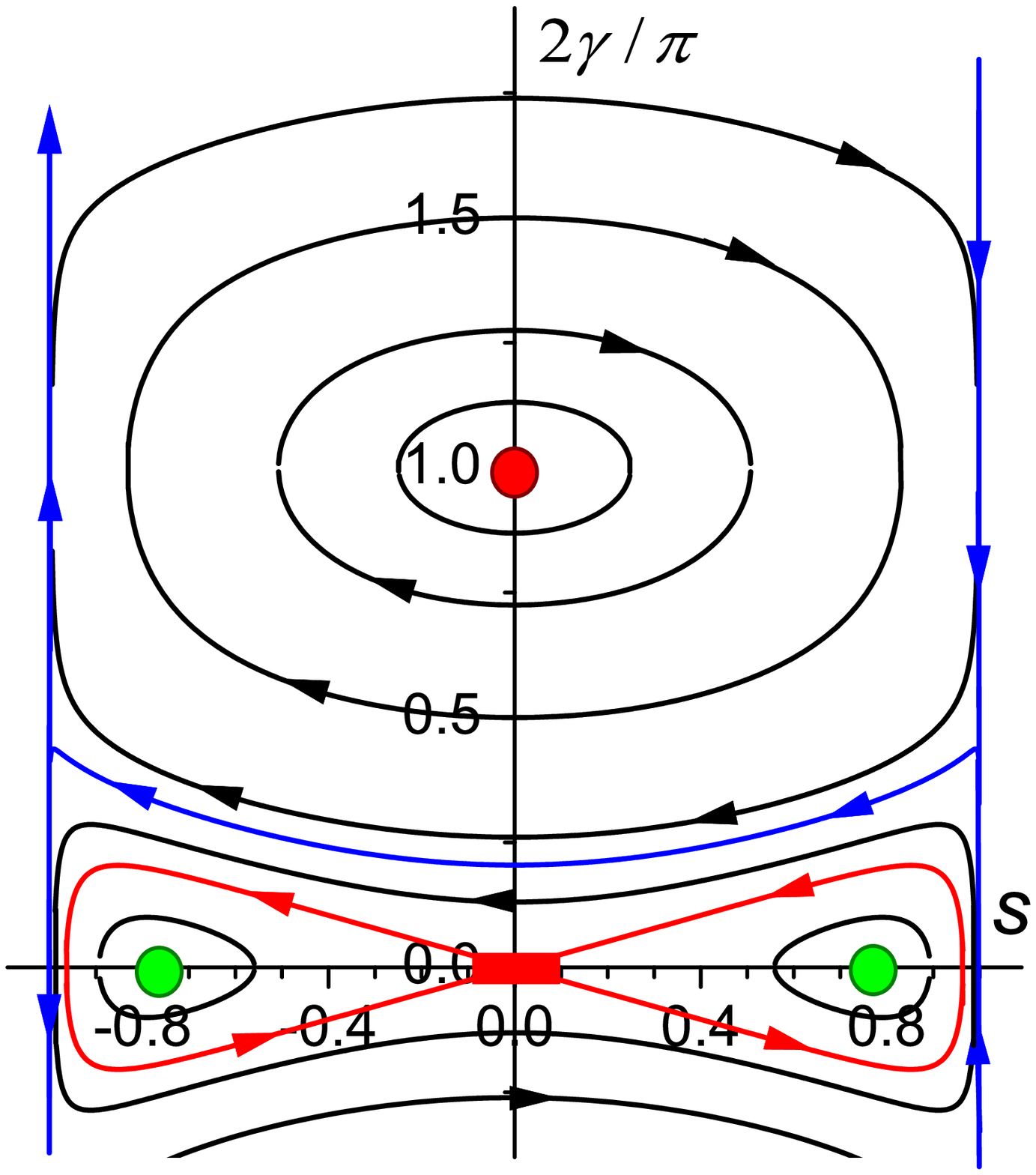}
 \label{2}}
 \subfigure[$\ \kappa =$0.5]{\includegraphics[width = 58mm]{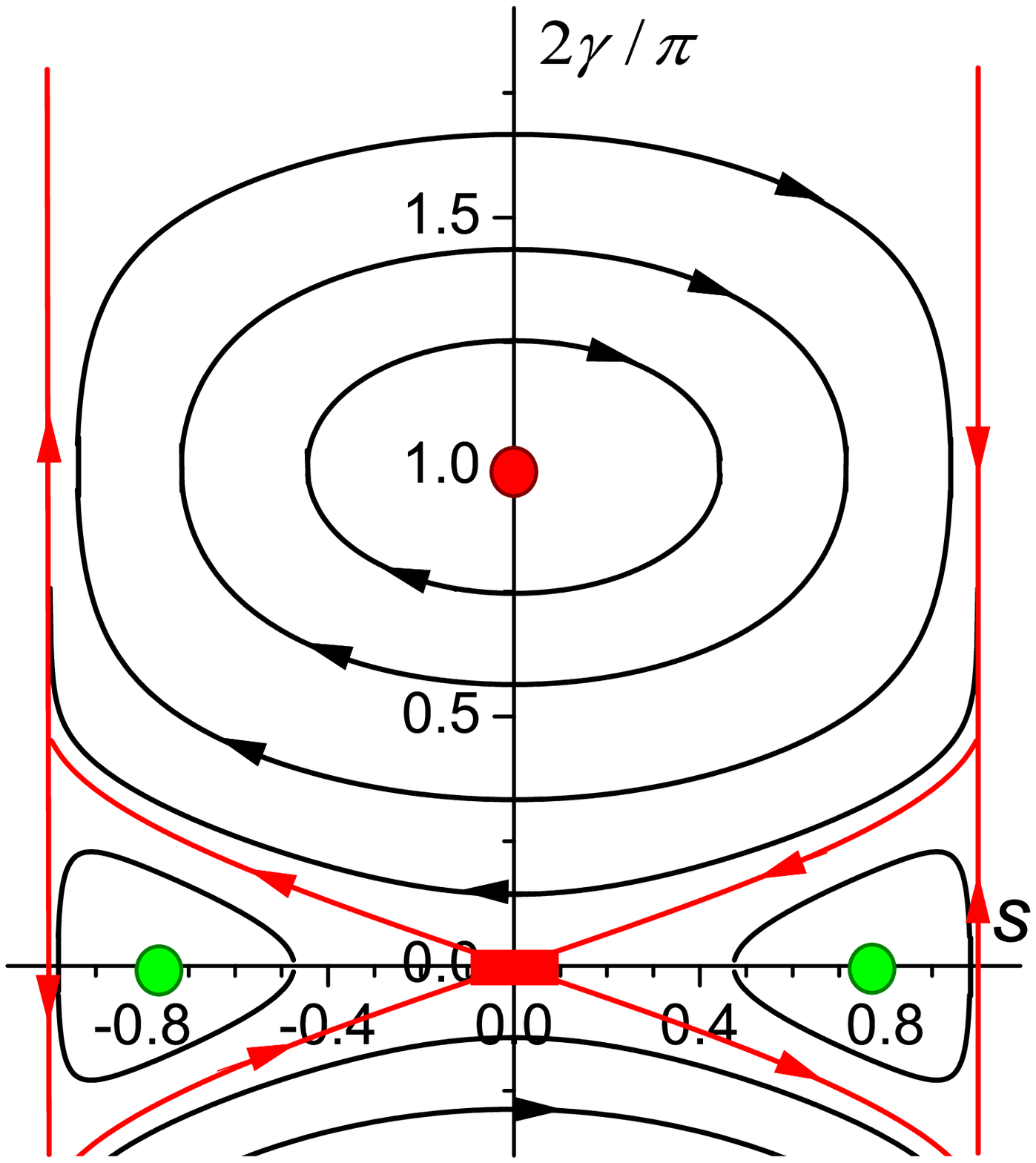}
 \label{3}}
 \subfigure[$ \ \kappa =$0.2]{\includegraphics[width = 58mm]{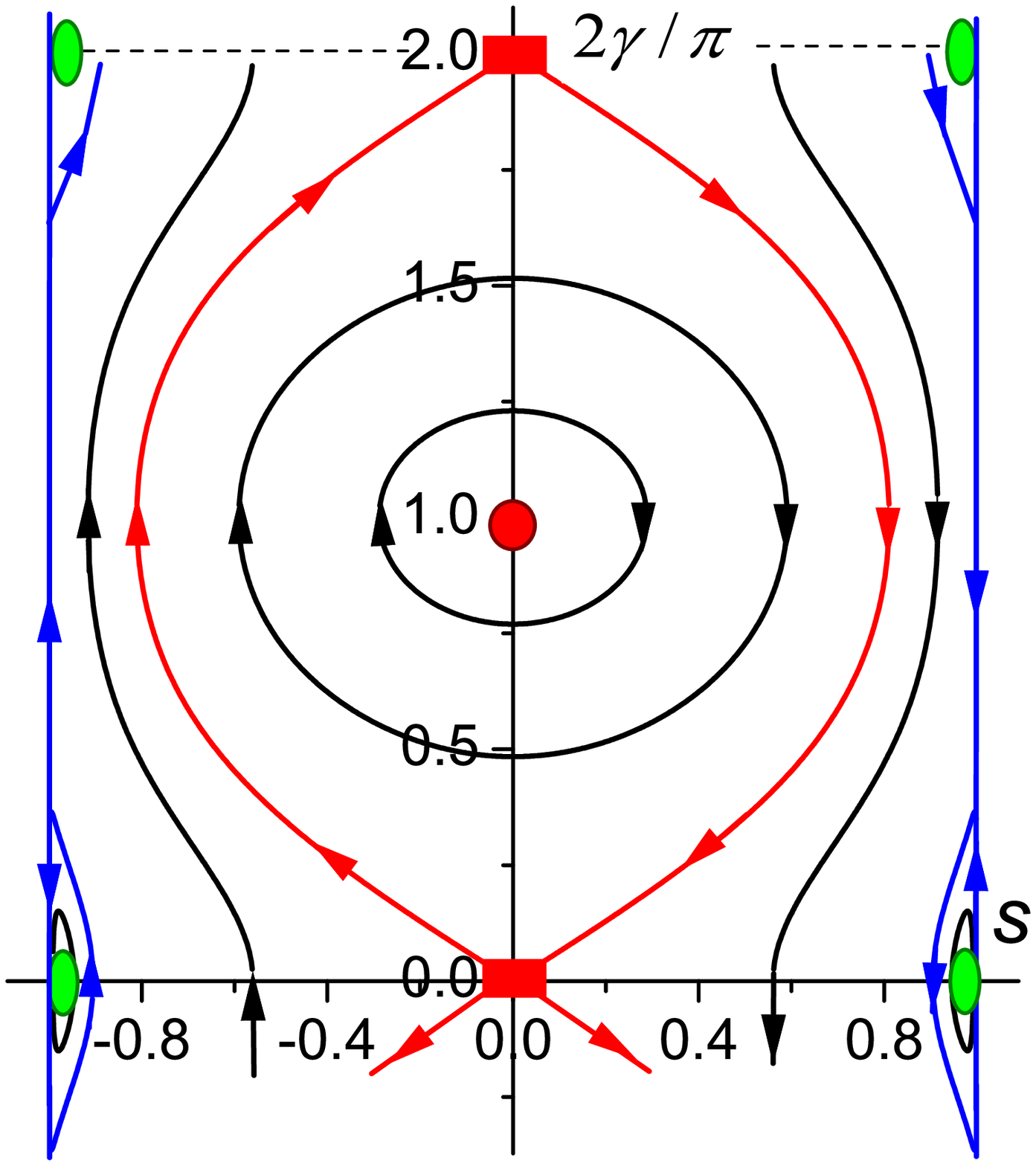}
 \label{4}}
 \caption{(color online) Phase plane representations for dissipation-free
non-linear longitudinal spin oscillations for different values of
the parameter $\kappa $; $\kappa =$0.6, 0.5 and 0.2 for panels (a),
(b), and (c), respectively. The green and red ellipses present the
minima and maxima, respectively; the standard saddle points are
depicted by red rectangles, while the standard separatrix
trajectories are drawn by red lines. The  singular   trajectories
with $s=\pm 1$ and the separatrix trajectories entering the
non-standard saddle points on these lines  $s=\pm 1$ are shown by
blue lines on the frames (a) and (c). For the critical value $\kappa
= 0.5$, all the separatrix trajectories and the singular
trajectories with  $s=\pm 1$ organize a common net; and on the
corresponding frame (b) all of them are presented by red lines. }
\label{fig2}
\end{figure}

\end{widetext}

\subsection{Damped longitudinal motion}
\label{ssec:damp}

Damping is a crucial ingredient for the dynamical switching between
different, but equivalent in energy, states. The high-frequency mode
of longitudinal oscillations have high-enough relative damping; as
was found from microscopic calculations,~\cite{[29]} the decrement
of longitudinal mode $\Gamma =\lambda \omega _l $, where $\lambda
\sim 0.2$. To account for the damping in the dynamic equations for
$s$ and $\gamma $, it is useful to consider a different
parametrization of the longitudinal dynamics. Let us now introduce a
unit vector, $\boldsymbol{\sigma}=\sigma _1 {\rm {\bf e}}_1 +\sigma
_2 {\rm {\bf e}}_2 +\sigma _3 {\rm {\bf e}}_3 $, with components
$\sigma _3 =s$, $\sigma _1 =\langle S_y^2 -S_x^2 \rangle ,$ and
$\sigma _2 =\langle S_1 S_2 +S_2 S_1 \rangle $. Being written
through $\boldsymbol{\sigma}$, the equation of motion takes the form
of the familiar Landau-Lifshitz equation
\begin{equation}
\label{eq9} \hbar \frac{\partial \boldsymbol{\sigma}}{\partial
t}=[\boldsymbol{\sigma}\times {\rm {\bf h}}_{\mathrm{eff}} ]+{\rm
{\bf R}}, \quad {\rm {\bf h}}_{\mathrm{eff}} =-\frac{\partial
W}{\partial \boldsymbol{\sigma}},
\end{equation}
where ${\rm {\bf h}}_{\mathrm{eff}} $ can be treated as an effective
field for longitudinal dynamics, and the relaxation term ${\rm {\bf
R}}$ is added. The equation of motion with ${\rm {\bf R}}=0$ is
fully equivalent to the Hamilton form of the equation found from
(\ref{eq7}), but the form of the dissipation is more straightforward
in unit-vector presentation. The choice of the damping term in a
standard equation for the motion of the transverse spin is still
under debate, see.~\cite{[15],[31]} But here the damping term can be
written in the simplest form, as in the original paper of Landau and
Lifshitz, ${\rm {\bf R}}=\lambda [{\rm {\bf h}}_{\mathrm{eff}}
-\boldsymbol{\sigma}({\rm {\bf h}}_{\mathrm{eff}}
\boldsymbol{\sigma})]$. The arguments are as follows: (i) this form
gives the correct value of the decrement of linear oscillations,
$\Gamma =\lambda \omega _l $; (ii) it is convenient for analysis,
because it keeps the condition $\boldsymbol{\sigma}^2=1$. Finally,
the equations of motion with the dissipation term of the
aforementioned form are:
\begin{equation}
\label{eq10}
\hbar \frac{ds}{dt}=-\frac{\partial W}{\partial \gamma }-\lambda
(1-s^2)\frac{\partial W}{\partial s},\;\hbar \frac{d\gamma
}{dt}=\frac{\partial W}{\partial s}-\frac{\lambda }{(1-s^2)}\frac{\partial
W}{\partial \gamma }.
\end{equation}
These equations describe the damped counterpart of the non-linear
longitudinal oscillations discussed in the previous subsection and
present as phase portraits on Fig. 2. The character of the motion at
not-too-large $\lambda $ can be qualitatively understood from energy
arguments. The trajectories of damped oscillations in any point of
the phase plane approximately follow the non-damped (described by
equation $W(s,\gamma )=\mathrm{const})$ ones, but cross them passing
from larger to smaller values of $W$, see Figs. 3 and 4. It happens
that for the case of interest, the dynamics is caused by the
time-dependent stimulus. An action of the stimulus on the system can
be described by adding the corresponding time-dependent interaction
energy $\Delta W$ to the system Hamiltonian, $W\to W(s,\gamma
)+\Delta W(s,\gamma ,t)$. Within this dynamical picture, $\Delta W$
produces an ``external force'' driving the system far from
equilibrium.

The analysis is essentially simplified for a pulse-like stimulus of
a short duration $\Delta t$ (much shorter than the period of motion,
$\omega _l \Delta t \ll 1)$. In this case, the role of the pulse is
reduced to the creation of some non-equilibrium state, which then
evolves as some damped nonlinear oscillations described by the
``free'' equations (\ref{eq10}) with $\Delta W=0$. The phase plane
method, which shows the behavior of the system for arbitrary initial
conditions, is the best tool for the description of such an
evolution.

\begin{figure}%[htbp]
\includegraphics[width=\figwidth]{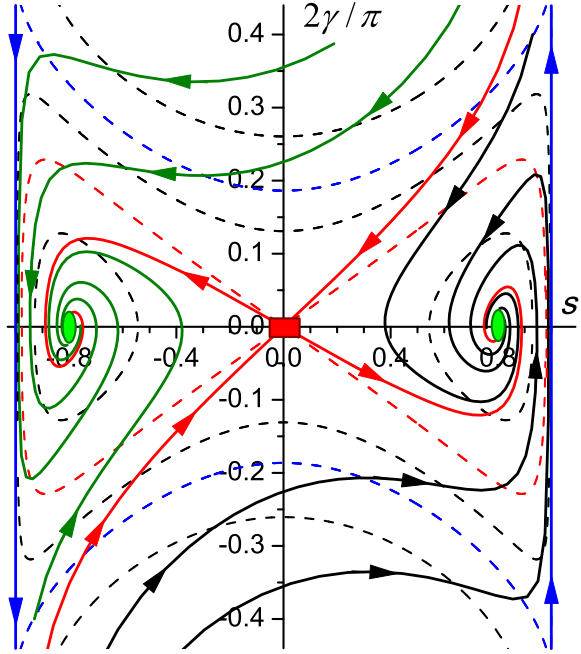}
\caption{\label{fig3}  (color online) Phase plane representation of
damped longitudinal spin oscillations for $\kappa =$0.6. Here and in
Fig. 4, the dashed lines (obtained analytically before) represent
the phase trajectories without dissipation, while the full lines are
trajectories for dissipation constant $\lambda =0.2$, found
numerically. The separatrix lines are drawn by red curves. }
\end{figure}

It is worth noting that the asymptotic behavior of the separatrix
trajectories at $\gamma ,s\to 0$, is important for this analysis,
and can be easily found analytically as
\begin{multline}
\label{eq11} \left( {\frac{\gamma }{s}} \right)_{\mathrm{separ}} =
R_{\mathrm{separ}} = \\  =\frac{1}{8\kappa }\left[ {\lambda
(1+3\kappa )+\sqrt {\lambda ^2(1+3\kappa )^2+16\kappa (1-\kappa )} }
\right]\quad .
\end{multline}

First let us start with the analysis for high-enough anisotropy. The
corresponding phase portrait is present in Fig.~3. The general
property of the phase plane is that the phase trajectories cannot
cross each other; they can only merge at the saddle points. Thus the
trajectories coming to different minima are stretched between two
separatrix lines entering the same saddle point from different
directions, as shown in Fig.~3. From this it follows that any
initial state with arbitrary non-equilibrium values of spin $s(+0)$,
but without deviation of $\gamma $ from its equilibrium value,
evolve to the state with the same sign of the spin as for $s(+0)$,
and no switching occurs. On the other hand, if the initial condition
is above the separatrix trajectory, entering the saddle point, the
evolution will move the system to the equivalent minimum with the
sign of the spin opposite to the initial one, $s(+0)$, realizing the
switching.

Figure 4 shows the phase plane for equations (\ref{eq10}) for the
more complicated case of low anisotropy, demonstrating possible
scenarios of the switching of the sign of the spin value during such
dynamics. Here the separatrix trajectories for the damped motion can
be monitored from their maxima, and the full picture of the behavior
can be understood only when including a few equivalent foci with
$\gamma =0,\,\pm \pi ,\, \pm 2\pi $, ect., with different, but
equivalent in energy, values of the spin, $s=\pm \bar {s}$, $\bar
{s}=\sqrt {1-\kappa ^2} $, and different saddle points, located at
$\gamma =0,\, \pm \pi ,\, \pm 2\pi $. As for small anisotropy, the
trajectories coming to different minima are located between two
branches of the separatrix lines, but now this ``separatrix
corridor'' is organized by separatrix lines entering different
saddle points. The switching phenomena is also possible, but the
process involves a few full turns of the variable $\gamma $.

The general regulation for any anisotropy can be formulated as
follows: for realizing spin switching, one needs to have the initial
deviation (reduction) of the spin value, and, simultaneously, a
non-zero deviation of the quadrupolar variable $\gamma $. To switch
the positive spin value to negative, one needs to start from the
states just above the separatrix line entering the saddle point from
positive values of $s$. The smaller the initial value of the spin,
the smaller value of $\gamma (0)$ would realize the switching. From
the asymptotic equation (\ref{eq11}), the corresponding ratio
$R_{\mathrm{separ}} =\gamma (0)/m(0)$ is smaller for small values of
$\lambda $; but even when $\lambda \to 0,$ it exceeds the value
$R_{\mathrm{separ}} (\lambda =0)=0.5\sqrt {(1-\kappa )/\kappa } $.
Thus, the switching could occur for non-zero values of $\kappa $.

\begin{widetext}

\begin{figure}
 \subfigure[]{\includegraphics[width = 51mm]{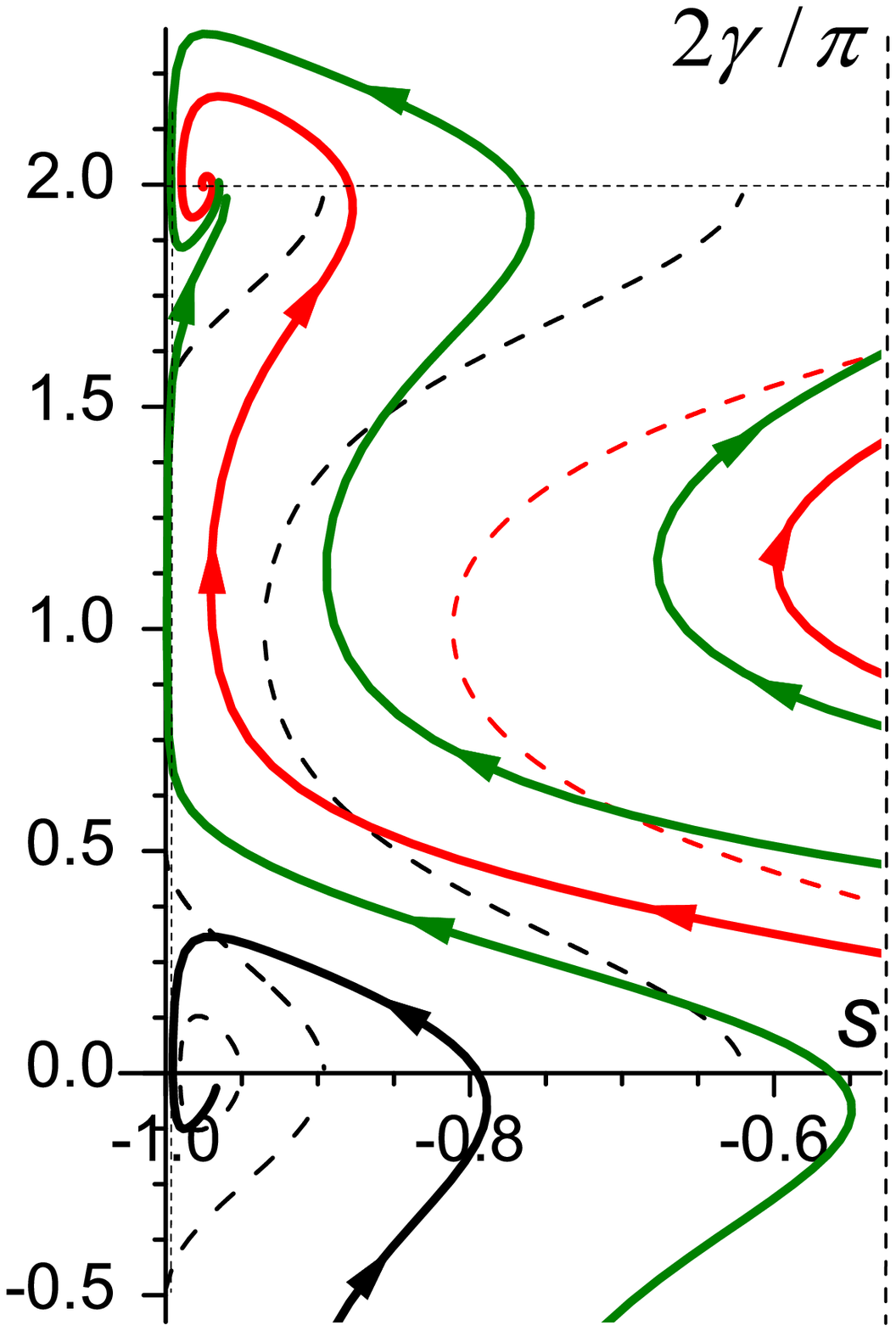}
 \label{6}}
 \subfigure[]{\includegraphics[width = 67mm]{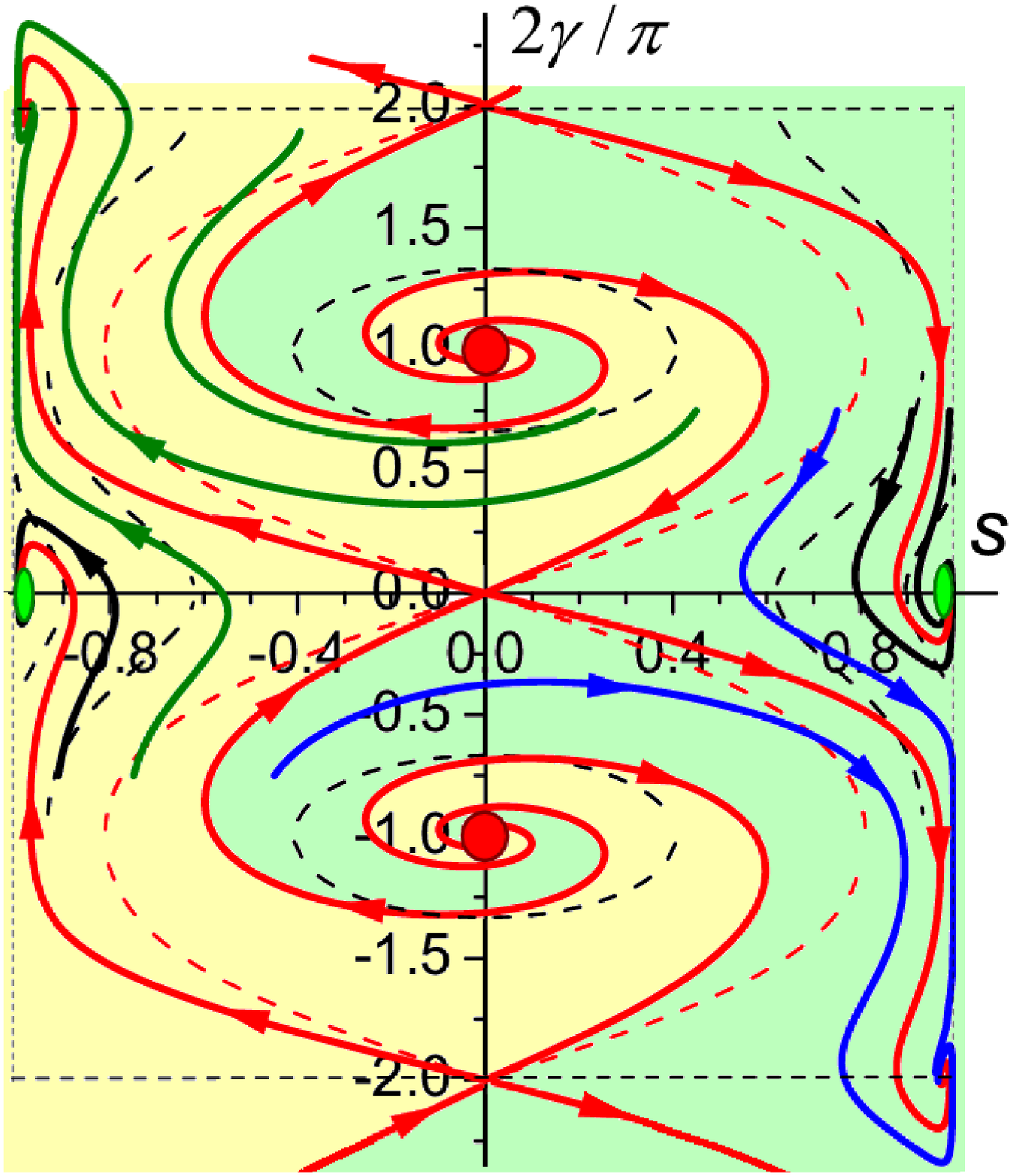}
 \label{7}}
 \subfigure[]{\includegraphics[width = 51mm]{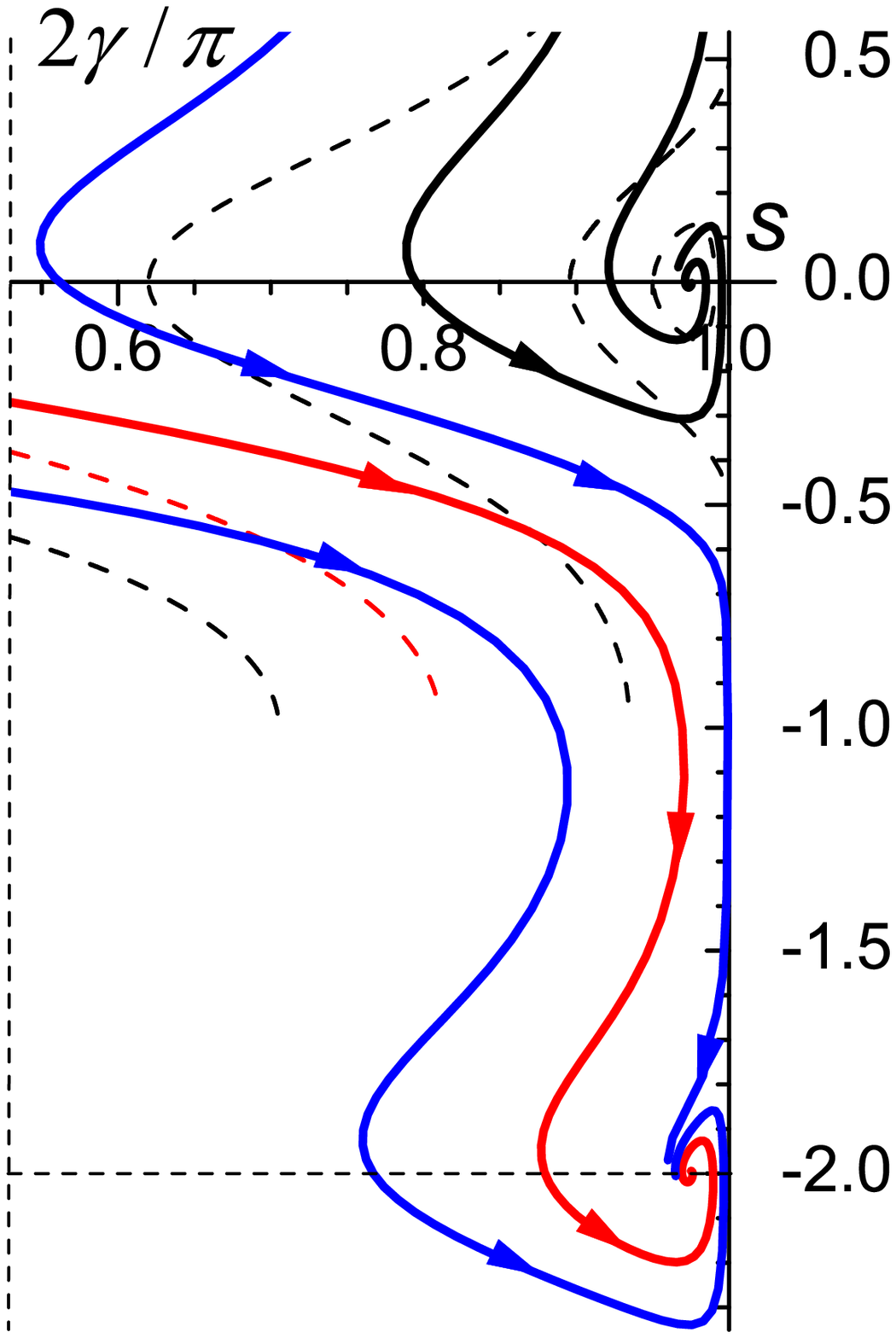}
 \label{8}}
 \caption{(color online) Phase plane for the damped spin evolutions for low
anisotropy, $\kappa =0.2$. The central frame shows the full diagram;
left and right panels demonstrate the details of the behavior near
the equilibrium values $s=-\bar{s}$ and $s=\bar{s}$, respectively.
On this frame, the regions colored by green and yellow correspond to
different basins of attraction with initial values leading to the
equilibrium states with $s=\bar{s}$ and $s=-\bar{s}$, respectively.}
\label{fig4}
\end{figure}

\end{widetext}

\section{INTERACTION OF THE LIGHT PULSE ON THE SPIN
SYSTEM: CREATION OF THE INITIAL STATE FOR SWITCHING.}
\label{sec:light}

Let us now consider the longitudinal spin evolution caused by a
specific stimulus: a femtosecond laser pulse. The reduction of the
spin to small values was observed in many experiments, and the only
non-trivial remaining question is: how can we create a deviation of
the quadrupolar variable $\gamma $ from its equilibrium value
$\gamma =0$. To find this, we now consider possible mechanisms of
light interaction with quadrupolar variables of non-Heisenberg
magnets.

The interaction of the spin system of magnetically-ordered media and
light is described by the Hamiltonian (as above, written per spin)
$\Delta W=\bar {\varepsilon }_{ij} v_0 E_i (t)E_j^\ast(t)  /16\pi $,
where $v_0 $ is the volume per spin, $E_i (t)$ is the time-dependent
amplitude of the light in the pulse, $\bar {\varepsilon }_{ij}
=d(\omega \varepsilon _{ij}^{(\mathrm{spin})} )/d\omega $,
$\varepsilon _{ij}^{(\mathrm{spin})} $ is the spin-dependent part of
the dielectric permittivity tensor, and $\omega $ is the frequency
of light. For the longitudinal dynamics considered here,
circularly-polarized light propagating along the $z$-axis acts on
the $z$-component of the spin via the standard inverse Faraday
effect, with the antisymmetric part of $\bar {\varepsilon }_{ij}
^{(a)}$ as, $\bar {\varepsilon }_{xy} ^{(a)}=-\bar {\varepsilon
}_{xy} ^{(a)}=s\alpha _\mathrm{F} $, giving an interaction of the
form
\begin{equation}
\label{eq12} \Delta W_{\mathrm{circular}} =s\frac{\alpha _\mathrm{F}
v_0 }{16\pi }\vert E_{\mathrm{circ}} \vert ^2\sigma ,
\end{equation}
where $E_{\mathrm{circ}} $ is the (complex) amplitude, $\sigma $
describes the pulse helicity: $\sigma =\pm 1$ for right-handed and
left-handed circularly polarized laser pulses. To describe
qualitatively the result of the action of the light pulse, let us
now assume that the pulse duration $\tau _{\mathrm{pulse}} $ is the
shortest time of the problem. If the pulse duration is shorter than
the period of spin oscillations, the real pulse shape can be
replaced by the Dirac delta function, $\vert E_{\mathrm{circ}} \vert
^2\to E_\mathrm{p} ^2\tau _{\mathrm{pulse}} \delta (t),$ where
$E_\mathrm{p} ^2=\int {\vert E_{\mathrm{circ}} \vert ^2dt} /\tau
_{\mathrm{pulse}} $ characterizes the pulse intensity. (Note that
this approximation is still qualitatively valid even for any
comparable values of $\tau _{\mathrm{pulse}} $ and $2\pi /\omega )$
Then, using equations (\ref{eq10}) one can find the effect produced
by the pulse. Within this approximation, the action of a pulse leads
to an instantaneous deviation of the variable $\gamma $ from its
equilibrium value, which then evolves following the non-perturbed
equations of motion (\ref{eq10}). Keeping in mind that before the
pulse action the system is in equilibrium, $s(-0)=\bar {s}$ and
$\gamma (-0)=0$, it is straightforward to find the values of these
variables $(s$ and $\gamma )$ after the action of the pulse, $s(+0)$
and $\gamma (+0)$. For our purposes, the non-equilibrium value of
the quadrupolar variable $\gamma $ is important:
\begin{equation}
\label{eq13} \gamma (+0)=-\frac{\alpha _\mathrm{F} }{16\pi \hbar
}E_\mathrm{p} ^2v_0 \tau _{\mathrm{pulse}} \sigma .
\end{equation}

The cumulative action of the circularly-polarized pulse, including
an essential reduction of the spin value (caused either by thermal
or non-thermal mechanisms) and the deviation of $\gamma $ described
by (\ref{eq13}) could lead to the evolution we are interested here,
switching the spin of the system. Note here that for standard spin
reduction the polarization of the light pulse is not
essential,~\cite{[32],[33],[34]} whereas the values of $\gamma (+0)$
are opposite for right- and left-handed circularly polarized pulses.
These features are characteristic of the effect described here. Note
the recent experiment where the role of circular polarization in
spin switching for GdFeCo alloy was mentioned, but the authors have
attributed it to magnetic circular dichroism.~\cite{[35]}

\section{CONCLUDING REMARKS.}
\label{sec:concl}

Let us now compare the approach developed in this article with
previous results on subpicosecond spin evolution. The first
experimental observation of demagnetization for ferromagnetic metals
under femtosecond laser pulses shows that the magnetic moment can be
quenched very fast to small values, much faster than one
picosecond.~\cite{[32],[33],[34]} These effects are associated with
a new domain of the physics of magnets,
\textit{femtomagnetism},~\cite{[36]} and its analysis is based on
the microscopic consideration of spins of atomic
electrons,~\cite{[37],[38]} or itinerant electrons.~\cite{[39]} Not
discussing this fairly promising and fruitful domain of magnetism,
note that, to the best of our knowledge, no effects of magnetization
reversal during this ``femtomagnetic stage'' has been reported in
the literature. For example, the subpicosecond quenching processes
for the ferromagnetic alloy GdFeCo are responsible for the creation
of a far-from-equilibrium state, but the evolution of this state,
giving the spin reversal, can be described within the standard set
of equations for the sublattice magnetizations.~\cite{[16]}

In contrast, here we propose some pathway to switch the sign of the
magnetic moment during extremely short times, of order of the
exchange time. It is shown here that the spin dynamics for magnets
with non-small single-ion anisotropy can lead to the switching of
the sign of the magnetic moment via the longitudinal evolution of
the spin modulus together with quadrupolar variables, i.e., quantum
expectation values of operators bilinear over the spin components
$S_x $ and $S_y $. It is worth to stress here that the ``restoring
force'' for this dynamics is the \emph{exchange interaction}, and
the characteristic time is the exchange time. On the other hand, to
realize this scenario, one needs to a have non-Heisenberg
interaction, e.g., single-ion anisotropy, which couples the spin
dipole and quadrupole variables.

Obviously, this effect is beyond the standard picture of spin
dynamics based on any closed set of equations for the spin dipolar
variables (i.e., the quantum expectation values linear on the spin
components) alone. Note that our approach based on the full set of
variables for the atomic spin is ``more macroscopic'' than the
``femtomagnetic'' approach,~\cite{[37],[38],[39]} dealing with
electronic states. To realize this type of switching, it is
necessary to have a significant coupling between dipolar and
quadrupolar spin variables, which is present in magnets with strong
single-ion anisotropy. Such anisotropy is known for the numerous
magnets based on anisotropic ions of transition elements such as
Ni$^{2+}$, Cr$^{2+}$, Fe$^{2+}$. As the classic example, note nickel
fluosilicate hexahydrate NiSiF$_6\cdot$6H$_2$O, with spin-one
Ni$^{2+}$ ions, coupled by isotropic ferromagnetic exchange
interaction and subject to high single-ion anisotropy. For this
compound, the strong effect of quantum spin reduction is known, with
its strength dependent on the pressure: the value of $K /J$ is
growing with the pressure $P$ resulting in the value $\langle S
\rangle =0.6$  at $P=6$ kbar and leading to the transition to the
non-magnetic state with $\langle S \rangle =0$ at $P\sim 10$
kbar.~\cite{Bar+NiSiF,Dyakonov+jetp}

A number of recent experiments  were done with rare-earth
transition-metals compounds.~\cite{[13],[14],[35],[40]}  Note here a
rich variety of non-linear spin dynamics observed for thin films of
the FeTb alloy under the action of femtosecond laser
pulses.~\cite{[40]} However, the theory developed here for simple
one-sublattice ferromagnet cannot be directly applied for the
description of such compounds. Ferromagnetic order with high
easy-plane anisotropy is present for many heavy rare-earth elements,
such as Tb and Dy at low temperatures.~\cite{REbook} This feature is
known both for bulk monocrystals,~\cite{REbook} and in thin layers
and superlattices, see Ref.~\cite{Dy_Ysuperlattices} and references
wherein. Strictly speaking, in our article only spin-one ions were
considered. Rare-earth metals have non-zero values of both spin and
orbital momentum, forming the total angular momentum of the ion, and
for their description the theory needs some modifications. However,
we believe that the effects of spin switching caused by quadrupolar
spin dynamics will be present as well for such magnets with high
values of atomic angular momentum.

The scenario proposed here includes inertial features, with the
evolution of an initial deviation from one equilibrium state to the
other, located far from the initial one. The initial deviations
should include both deviation of the magnetization and of the
quadrupolar variable, $\gamma$. Thus the effect is based on standard
magnetization reduction, but it is helicity dependent as well. The
necessary initial deviations can be created by a light pulse of
circular polarization and the possibility of switching depends on
the connection of the initial spin direction and the pulse helicity.
The possible materials should satisfy a number of general
conditions: they should be very susceptible to magnetization
quenching, which is typical for many materials, as well as a
sizeable Faraday effect, and they should also have spin one and a
high enough easy-plane anisotropy.

This work is partly supported by the Presidium of the National
Academy of Sciences of Ukraine via projects no.VTs/157 (EGG) and
No.0113U001823 (BAI) and by the grants from State Foundation of
Fundamental Research of Ukraine No. F33.2/002 (VIB and YuAF) and No.
F53.2/045 (BAI). FN acknowledges partial support from the ARO, RIKEN
iTHES Project, MURI Center for Dynamic Magneto-Optics, JSPS-RFBR
Contract No. 12-02-92100, Grant-in-Aid for Scientific Research (S),
MEXT Kakenhi on Quantum Cybernetics, and Funding Program for
Innovative R\&D on S\&T (FIRST).

%%%%%%%%%%%%%%%%%%%%%%%%%
\end{document}